\documentclass{article}
\usepackage{cite}
\usepackage{graphicx}
\usepackage{dcolumn}

\begin{document}

\date{}
\title{Comment on ``An appropriate approach to pricing european-style options with
the Adomian decomposition method''}
\author{Francisco M. Fern\'{a}ndez\thanks{%
fernande@quimica.unlp.edu.ar} \\
%EndAName
INIFTA, DQT, Sucursal 4, C. C. 16, \\
1900 La Plata, Argentina}
\maketitle

\begin{abstract}
We show that the Adomian decomposition method proposed by Ke et al [ANZIAM
J. \textbf{59} (2018) 349] is just the Taylor series approach in disguise.
The latter approach is simpler, more straightforward and yields a recurrence
relation free from integrals.
\end{abstract}

In a recent paper published in this journal Ke et al\cite{KGZ18}
(KGZ from now on) applied the so called Adomian decomposition
method (ADM) to the Black-Scholes model for the pricing of
European options. In order to avoid the problem of
differentiability of the solution to the partial differential
equation at origin they moved the discontinuity at infinity by
means of a suitable change of variables. The purpose of this
Comment is the analysis of the implementation of the ADM proposed
by KGZ.

Our starting point is the partial differential equation derived by KGZ after
the change of variables:
\begin{equation}
\partial _{z}\left[ zu(y,z)\right] =2\partial _{y}^{2}u(y,z)+y\partial
_{y}u(y,z)+2\left( k_{1}-1\right) z\partial _{y}u(y,z)-2k_{2}z^{2}u(y,z),
\label{eq:PDE}
\end{equation}
where $\partial _{z}$ and $\partial _{y}$ denote partial derivatives with
respect to $z$ and $y$, respectively. Here, we expand the solution $u(y,z)$
in a Taylor series about $z=0$:
\begin{equation}
u(y,z)=\sum_{j=0}^{\infty }f_{j}(y)z^{j},\;f_{j}(y)=\frac{1}{j!}\left.
\partial _{z}^{j}u(y,z)\right| _{z=0}.  \label{eq:u_Taylor}
\end{equation}

If we insert the expansion (\ref{eq:u_Taylor}) into the differential
equation (\ref{eq:PDE}) we obtain a recurrence relation for the functions $%
f_{j}(y)$:
\begin{eqnarray}
f_{n}(y) &=&\frac{1}{n+1}\left[ 2\partial _{y}^{2}f_{n}(y)+y\partial
_{y}f_{n}(y)+2\left( k_{1}-1\right) \partial
_{y}f_{n-1}(y)-2k_{2}f_{n-2}(y)\right] ,  \nonumber \\
n &=&0,1,\ldots ,\;f_{-1}=f_{-2}=0.  \label{eq:f_n_rec_rel}
\end{eqnarray}
It is not difficult to verify that this recurrence relation enables one to
obtain all the functions $u_{j}(y,z)=f_{j}(y)z^{j}$ derived by KGZ by means
of the ADM through a lengthier and more cumbersome procedure. We can show
the connection between the Taylor and ADM solutions in an even clearer way.

If we multiply equation (\ref{eq:f_n_rec_rel}) by $z^{n}$ and take into
account that $z^{n}/(n+1)=z^{-1}\int_{0}^{z}z^{n}\,dz$ we obtain
\begin{eqnarray}
f_{n}(y)z^{n} &=&\frac{1}{z}\int_{0}^{z}\left[ 2\partial
_{y}^{2}f_{n}(y)z^{n}+y\partial _{y}f_{n}(y)z^{n}+2\left( k_{1}-1\right)
z\partial _{y}f_{n-1}(y)z^{n-1}\right. \,  \nonumber \\
&&\left. -2k_{2}z^{2}f_{n-2}(y)z^{n-2}\right] \,dz,  \label{eq:u_n_rec_rel}
\end{eqnarray}
that is exactly the equation derived by KGZ by means of the ADM if $%
u_{j}(y,z)=f_{j}(y)z^{j}$.

It is clear that the implementation of the ADM proposed by KGZ is simply the
Taylor-series approach in disguise. The latter is by far simpler and more
straightforward and yields the expression (\ref{eq:f_n_rec_rel}) that is
free from the integral over $z$. It is worth adding that the solutions
provided by this particular form of the ADM are valid for sufficiently small
values of the expansion variable $z$. This is a well known limitation of any
approach based on a power-series expansion.


\begin{thebibliography}{9}
\bibitem{KGZ18}  Z. Ke, J. Goard, and S-P. Zhu, ``An appropriate approach to
pricing european-style options with the Adomian decomposition method'',
ANZIAM J. \textbf{59}, 349-369 (2018).
\end{thebibliography}
\end{document}